\documentclass[twocolumn,showpacs,preprintnumbers,amsmath,amssymb,prl]{revtex4}

\usepackage{graphicx,mathrsfs}
\usepackage{dcolumn}
\usepackage{bm}

\begin{document}

\newcommand{\bra}[1]{\ensuremath{\left< #1 \right|}}
\newcommand{\ket}[1]{\ensuremath{\left| #1 \right>}}
\newcommand{\ud}{\mathrm{d}}
\newcommand{\E}{\mathscr{E}}
\newcommand{\kbf}{\mathbf{k}}


\title{Recurrent fourth-order interference dips and peaks with a comb-like two-photon entangled state}

\author{Alessandro Zavatta}

\author{Silvia Viciani}

\author{Marco Bellini}

\altaffiliation[Also at ]{
LENS and 
INFM, Florence, Italy} \email{bellini@inoa.it}
\affiliation{Istituto Nazionale di Ottica Applicata (INOA),\\L.go
E. Fermi, 6, I-50125, Florence, Italy}

\date{\today}

\begin{abstract}
We demonstrate full selective control over the constructive or destructive character of
fourth-order recurrent interferences in a modified version of a Hong-Ou-Mandel interferometer
using comb-like two-photon states. The comb spectral/temporal structure is obtained by
inserting an etalon cavity in the signal path of an entangled photon pair obtained by pulsed
spontaneous parametric down-conversion. Both a simple qualitative discussion and a complete
theoretical derivation are used to explain and analyze the experimental data.
\end{abstract}


\maketitle

In the original scheme by Hong, Ou, and Mandel (HOM) \cite{hong87} two single photons from a
spontaneous parametric down-conversion (SPDC) pair are sent to two input ports of a
beam-splitter (BS) and coincidences are observed between the detection events on two detectors
placed at the BS output ports. While varying the relative phase delay between the two beams,
one can observe a dip in the coincidence signal when such a delay is shorter than the
coherence time of the downconverted photons. This fourth-order interference effect takes place
when two photons in the same mode arrive simultaneously at BS and can also be observed when
they are emitted independently by a single-photon device \cite{santori02}. However, when an
entangled two-photon state is considered, it is the indistinguishability between two
two-photon amplitudes leading to the same detector "firing scheme" that gives rise to
fourth-order interferences, and this may happen even if the two single photons arrive at BS at
two different times (with a delay which can be much longer than their coherence time) and
follow two distinguishable optical paths to reach the detectors. The interference will be
either destructive or constructive, resulting in a dip or a peak in the coincidence signal,
depending on the phase difference between these two two-photon amplitudes, as first observed
by the group of Shih~\cite{pittman96,strekalov98,kim99,burlakov01}.

\begin{figure}
\center
\includegraphics[width=8.0 cm]{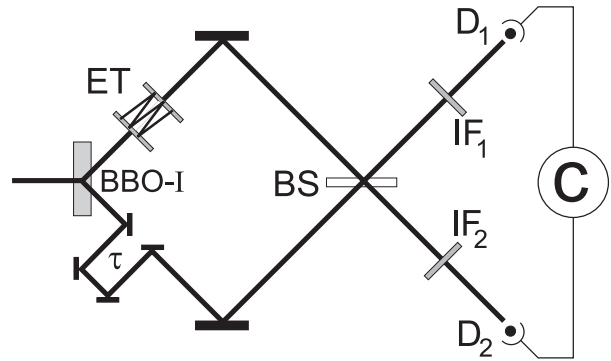}
\caption{Schematic view of the experimental set-up. See the text
for details. \label{fig_HOMexp}}
\end{figure}

Here, by using comb-like entangled states in a modified version of the HOM setup, we are able
to force the concurrent contribution of both (HOM- and Shih-type) kinds of interferences to
the generation of finely controllable dips or peaks in a recurrent pattern. A variable delay
line is inserted in the idler path while an etalon cavity is placed in the other beam path and
modifies the temporal structure of the signal photon wavepacket. The two photon wavepackets
are then mixed at BS and coincidence events between detectors D$_1$ and D$_2$ at its output
ports are measured while scanning the delay line (see Fig.\ref{fig_HOMexp} for a simplified
scheme of the experiment). We demonstrate a full control over the constructive or destructive
character of the recurrent fourth-order interferences by means of a tuning of the etalon
cavity in an easy and simply predictable way. This type of measurement has been recently
discussed in a theoretical paper by Pe\v rina~\cite{perina03} and a somewhat related
experiment with so-called mode-locked two-photon states has also been recently reported by Lu
et al.~\cite{lu03}.

The etalon cavity is characterized by a comb-like spectral
transmission function, with peaks equally spaced by the
free-spectral-range (FSR), corresponding to the inverse of the
cavity round-trip time $T$ ($1/FSR=T=2d/c$, where $d$ is the
mirror separation and $c$ the speed of light). If the bandwidth of
the down-converted photons is larger than the cavity FSR, several
transmission modes of the resonator are simultaneously excited by
the incoming field, and the output signal spectrum acquires a
comb-like structure made of different peaks. It is also
interesting to look at the effects of the etalon in the temporal
domain: if the temporal coherence of the signal photon is much
shorter than the etalon round-trip time, a signal photon
wavepacket cannot interfere with its leading edge having been
reflected twice by the mirrors inside the cavity. The result is a
train of isolated coherent wavepackets of exponentially decreasing
amplitude leaving the etalon equally spaced by the cavity
round-trip time $T$. In general, the phase delay between
consecutive pulses can be expressed as $\Delta \varphi=2 \pi n
+\delta \varphi$, where $n$ is a positive integer and only the
term $\delta \varphi$ plays a physically significant role. Note
that the phase delay $\delta \varphi$ can be easily varied in a
controllable way by slightly rotating the etalon along an axis
perpendicular to the beam propagation direction, thus changing the
optical path of the beam inside the cavity.

\begin{figure}
\center
\includegraphics[width=8.5 cm]{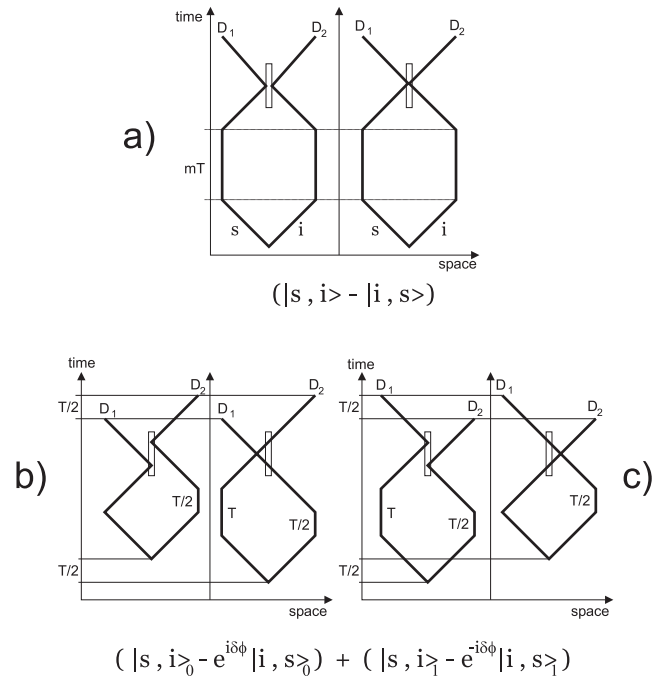}
\caption{a): Feynman-like diagrams for the intuitive case where the idler delay is set equal
to an integer number of etalon round-trips and photons arrive simultaneously at BS (HOM case).
b) and c): The two pairs of interfering amplitudes for an idler delay corresponding to half a
cavity round-trip; the two photons reach BS at different times (Shih case). \label{fig_HOM1}}
\end{figure}

Now let us consider what happens when a single signal photon enters the etalon: in some cases
it may be transmitted by both cavity mirrors and arrive to BS with just a small delay
connected with the crossing of the mirror material; this is equivalent to the HOM case, and a
dip in the coincidence rate is expected when the delay in the idler path is properly adjusted.
Let us define this delay as $\tau=0$. Interference arises from the indistinguishability of the
two possibilities for photons from the two paths of being both either transmitted or reflected
by BS. If the mirror reflectivity is high enough, it is however more probable for the pulse to
oscillate a few times inside the cavity before going out towards BS after a delay $mT$
corresponding to an integer number $m$ of round-trips. If one just thinks of interference as
arising from the temporal overlap of two one-photon wavepackets on the beamsplitter, then one
might expect to see partial revivals of the interference dip also for idler path delays
$\tau_m=mT$. Again, from a coincident detection event on the two detectors one cannot tell,
not even in principle, if the photons have been both reflected or both transmitted by BS (see
Fig.\ref{fig_HOM1}a)). For each $\tau_m$, both these alternatives contribute with the same
phase term, which thus factors out, leaving with the usual HOM state and with the appearance
of a dip. According to this intuitive approach, only dips for delays $\tau_m$ equal to integer
multiples of the cavity round-trip time $T$ are then to be expected. However, other
indistinguishable alternatives leading to the same detector firing schemes (not necessarily
coincidences!) are possible, and may thus lead to additional quantum interferences.

If the idler delay line is set such that $\tau=T/2$ two firing
configurations are possible, both of them allowing for two
indistinguishable realizations represented by the Feynman-like
diagrams of Fig.\ref{fig_HOM1}b) and c). In the first one (see
Fig.\ref{fig_HOM1}b)), detector D$_1$ fires ahead of detector
D$_2$ by a time $T/2$; this detection event can be realized either
by letting the signal photon pass without reflections through the
etalon and having both photons reflected by BS, or with one signal
round-trip in the etalon and two transmissions through BS. In the
second firing scheme, detector D$_2$ clicks ahead of detector
D$_1$ by the same amount of time, as shown in
Fig.\ref{fig_HOM1}c)): it can be realized either by a single
signal round-trip through the etalon and two reflections on BS, or
by transmission through the etalon and transmission of both
photons through BS. In both these firing schemes the two
interfering probability amplitudes differ by a phase term $\delta
\varphi$ corresponding to one round-trip in the cavity. If $\delta
\varphi$ is set to zero by an appropriate tilt of the etalon, the
phase term cancels and a dip appears, but if $\delta \varphi=\pi$
then the dip becomes a peak while, for $\delta \varphi=\pi/2$, the
coincidence rate flattens out.
Note that the two alternatives in both the above firing schemes
imply a different time of birth of the photon pair inside the
crystal, and are truly indistinguishable only if the difference
between the times of generation ($T/2$ in this case) is smaller
than the pump coherence time. Although this is always true for a
CW pump, it may constitute an important limitation to the
visibility of interferences in the case of a pulsed pump with a
short coherence time.

\begin{figure}
\center
\includegraphics[width=8.5 cm]{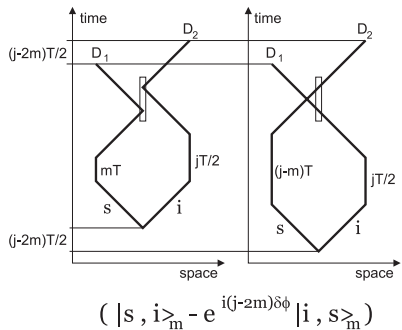}
\caption{General diagram for one of the ($j+1$) terms contributing
to quantum interferences in the coincidence counts when the idler
delay is set to an integer number of half etalon round-trips:
$\tau_j=jT/2$.\label{fig_HOMgen}}
\end{figure}

The above discussion can be generalized to the case where the
idler photon delay is an integer multiple $j$ of half the
round-trip time (i.e. $\tau_j=j T/2$). If $m$ is the number of
round-trips of the signal photon inside the etalon, one easily
finds that, for each $j$, there are $j+1$ (corresponding to $m$
going from 0 to $j$) firing schemes possible, each allowing for
two indistinguishable realizations. Each firing scheme (denoted by
$m$ and illustrated in Fig.\ref{fig_HOMgen}) is the result of two
alternative probability amplitudes, one involving $m$ and the
other $(j-m)$ signal round-trips, which present a phase difference
of $(j-2m)\delta \varphi$.  Note that this phase difference is an
even or odd multiple of $\delta \varphi$ depending on the parity
of $j$.

\begin{figure}
\center
\includegraphics*[width=8.0 cm]{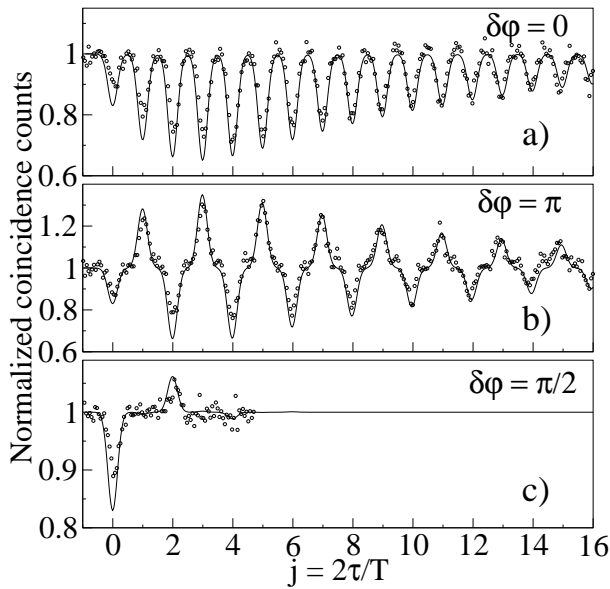}
\caption{ \label{exp-sim} Experimental results and calculated
curves for the coincidence rates as a function of idler delay for
three settings of the cavity inter-pulse phase delay.}
\end{figure}

In order to give a simple expression for the final comb-like
two-photon state and gain some insight into the terms contributing
to these interference effects, we may make some crude
approximations: we may assume that the pump is monochromatic (its
coherence length is infinite) and that the reflectivity of the
etalon mirrors is very high, so that we may neglect the amplitude
decay between successive wavepackets exiting the cavity. The
two-photon state at a given idler delay $\tau_j=j T/2$, can then
be written as:
\begin{equation}
\ket{\psi_{j}} \propto \sum_{m=0}^{j}\left(\ket{s,i}_m -
e^{i(j-2m) \delta \varphi} \ket{i,s}_m\right) \label{eq1}
\end{equation}
where the $\ket{s,i}_m$ ($\ket{i,s}_m$) term corresponds to the
probability amplitude of a signal photon being detected at D$_1$
(D$_2$) and an idler photon at D$_2$ (D$_1$) for a given firing
scheme, and corresponds to the first (second) Feynman-like diagram
of Fig.\ref{fig_HOMgen}.
For $\delta \varphi=0$, all the terms in eq.\ref{eq1} contribute with a null phase, so that a
sequence of dips is expected at idler delays corresponding to all $\tau_j=jT/2$ (see
Fig.\ref{exp-sim}a)).

For even $j$, one of the firing schemes (the one with $m=j/2$) always corresponds to the HOM
case described earlier and depicted in Fig.\ref{fig_HOM1}a). For $j=0$ this is the only
possible contribution, so that for $\tau=0$ a dip is always present, independent of $\delta
\varphi$. For $j>0$ however, there are additional (Shih-type) contributions which, depending
on $\delta \varphi$, may be either constructive or destructive, so that the expected dip may,
in some cases, flatten out or become a peak. Consider for example the case of $j=2$ (i.e. the
idler delay is set at the cavity round-trip time $T$): as shown above, three contributions are
present, one ($m=1$, HOM-type) for which the phase term is zero and always contributes as a
dip, and two others ($m=0,2$ Shih-type) with phase terms equal to $\pm 2 \delta \varphi$ which
can be adjusted by rotating the etalon. If $\delta \varphi=\pi/2$ (see Fig.\ref{exp-sim}c))
they both contribute as peaks, so that the final coincidence rate should show a small peak in
this position. With this setting of the phase, it is also clear that all the delays
corresponding to an odd $j$ (pure Shih-type interferences) will give a flat coincidence rate,
being determined by contributions with phase terms which are odd multiples of $\pi /2$. For
the same reason, all these delay positions will change to peaks for $\delta \varphi=\pi$
(Fig.\ref{exp-sim}b)).

In this simplified situation, the behavior of the coincidences for
different idler delays and for different settings of the etalon
inter-pulse phase can be simply obtained by extending the above
reasoning. Although useful for an intuitive understanding of the
process and for a qualitative prediction of the experimental
results, the above description is however too crude for a direct
comparison of the measured data with calculations. A more refined
and complete approach has then to be used, which takes into
account all the parameters of the real experimental situation,
like the finite coherence of the pump, the reflectivity of the
etalon mirrors, the presence of spectral filters in the photon
paths, and the natural spectral width of the SPDC two-photon
state.

For low time-resolution detectors, the coincidence rate is
obtained by a time integration of the second-order correlation
function
\begin{equation*}
R_c = \int \ud t_1 \ud t_2 \bra{\psi}\hat E_1^{(-)}(t_1)\hat E_2^{(-)}(t_2)
\hat E_2^{(+)}(t_2)\hat E_1^{(+)}(t_1)\ket{\psi}.
\end{equation*}
The two-photon state generated by spontaneous parametric down-conversion
in a $\chi^{(2)}$ nonlinear crystal is given by~\cite{ou97,keller97}
\begin{equation*}
\ket{\psi} =\int \ud\omega_s \ud\omega_i
\phi(\omega_s,\omega_i) \ket{\omega_s}_s\ket{\omega_i}_i
\end{equation*}
where $\ket{\omega}_s$ and $\ket{\omega}_i$ are the single photon
states for the signal and idler mode with frequency $\omega$, and
the probability amplitude
\begin{equation*}
\phi(\omega_s,\omega_i) = \alpha \E^{(+)}_p(\omega_s+\omega_i)
\frac{\sin(\Delta kL/2)}{\Delta kL/2}e^{-i\Delta kL/2}
\end{equation*}
for the two-photon state is the product of the pump envelope
$\E^{(+)}_p(\omega)$ with the phase-matching function, which
depends on the crystal phase-mismatch $\Delta k$ and on the
crystal length $L$. The term $\alpha$ includes all the constants
and the slowly varying terms. The field operators at the detectors
are given by
\begin{eqnarray*}
E_{1,2}^{(+)}(t) &=&\frac{1}{\sqrt{2}}\int \ud \omega f(\omega)
\Big(\hat a_s(\omega)f_e(\omega)\pm \hat
a_i(\omega)e^{i\omega\tau} \Big) e^{i\omega t}
\end{eqnarray*}
where $\hat a_s(\omega)$, and $\hat a_i(\omega)$ are the
annihilation operators for the signal and idler fields with
frequency $\omega$ and $f(\omega)$ is the transmission function
(assumed of Gaussian shape) of the interference filter placed in
front of each detector. The etalon inserted in the idler beam is
described by \cite{siegman}
\begin{equation*}
f_e(\omega) = \frac{(1-R)e^{i  \omega d/c}}{1-Re^{ 2 i  \omega d \cos\theta/c}}
\end{equation*}
where $d$ is the etalon mirror separation, $R$ the mirror
reflectivity and $\theta$ the internal incidence angle. The
coincidence rate is thus proportional to
\begin{eqnarray}
&&R_c(\tau)=\frac{1}{4} \int \ud \omega_i \ud \omega_s F(\omega_s)
F(\omega_i) \Big\{|\phi(\omega_s,\omega_i)|^2 |f_e(\omega_s)|^2
\nonumber
\\ &&- \phi(\omega_s,\omega_i)\phi^*(\omega_i,\omega_s)
f_e(\omega_s)f_e^*(\omega_i)e^{-i(\omega_s-\omega_i)\tau}\Big\}\label{eq:Rc}
\end{eqnarray}
with $F(\omega)=|f(\omega)|^2$. When the etalon and interference
filters are removed, the above expression is equivalent to that
for the single HOM dip given in~\cite{grice97}.

The comb-like two-photon entangled state is experimentally generated by Type-I SPDC in a 3-mm
long BBO crystal. The crystal is slightly tilted from the collinear condition in order to get
degenerate parametric emission along a cone when pumped by the second harmonic of 1.4~ps-long
pulses from a mode-locked Ti:sapphire laser operating around 786~nm. The signal and idler
photons are then selected by two apertures and sent to the two input ports of a 50\%
beam-splitter (BS), as schematically shown in Fig.~\ref{fig_HOMexp}. The air-spaced etalon
placed on the path of the signal beam is made of a pair of plane and parallel facing mirrors
with a reflectivity of 90\% for a wavelength of 786 nm, placed at a distance of 100 $\mu$m.
The cavity FSR is 1500 GHz (round-trip time $T$~=~0.67~ps), corresponding to a width of about
3.1~nm at the central wavelength of the signal photon. The bandwidth of the downconverted
photons is much larger than the etalon FSR and is anyway limited by the 10~nm-wide
interference filters (IFs) placed in front of the detectors. A motorized translation stage,
placed on the path of the idler beam, is used to finely control the delay $\tau$ between the
signal and idler photons. The beams emerging from the output ports of the BS are collected by
means of 25~mm focal-length graded-index lenses, and detected by two single-photon counting
modules, D$_1$ and D$_2$ (SPCM Perkin-Elmer AQR-12). Figure~\ref{exp-sim} shows the measured
D$_1$-D$_2$ coincidence counts as a function of the idler delay for different settings of the
inter-pulse phase $\delta \varphi$. Note that our 2~ns gate width also counts all the
"delayed" coincidences depicted in Fig.\ref{fig_HOMgen} as valid events. The condition $\delta
\varphi = 0$ is achieved by tilting the etalon so that one of its transmission maxima
coincides with the interference filters' transmission peak. The phase delays $\delta \varphi =
\pi$ and $\delta \varphi = \pi / 2$ are instead obtained by slight rotations of the cavity
such that its comb-like spectrum is frequency shifted by FSR/2 and FSR/4, respectively. The
solid curves in Fig.~\ref{exp-sim} show the result of the theoretical calculations according
to eq.~(\ref{eq:Rc}) without adjustable parameters. The agreement between the experimental and
theoretical data is excellent and they both clearly confirm our previous qualitative
discussion. The demonstration of an accurate control over the character of the individual
interferences in the array produced by the comb-like entangled state may prove important for
the implementation of new schemes in quantum state engineering and information processing.

This work has been supported by the Italian Ministry of University
and Scientific Research (MIUR). The authors would also like to
thank the Physics Dept. of the University of Florence for the kind
hospitality and F. Marin for useful discussions.

\bibliography{nl_bib}

\end{document}